# Levitation in an "almost" electrostatic field

E. N. Miranda

CRICYT – CONICET

5500 – Mendoza, Argentina

and

Dpto. de Física

Universidad Nacional de San Luis

5700 – San Luis, Argentina

**Abstract:** It is well known that a charged particle cannot be in stable equilibrium in a purely electrostatic field. The situation is different in a magnetostatic field; consequently, magnetic levitation is possible while electrostatic levitation is not. In this paper, motivated by an analogy with a mechanical system, we show that the addition of a small oscillating electrical field to an otherwise electrostatic configuration leads to the stabilisation of unstable equilibrium points. Therefore, levitation becomes possible in an "almost electrostatic" field.

# Levitación en un campo "casi" electrostático

**E. N. Miranda**

CRICYT – CONICET

5500 – Mendoza, Argentina

y

Dpto. de Física

Universidad Nacional de San Luis

5700 – San Luis, Argentina

**Resumen:** Se sabe que una partícula cargada no puede estar en equilibrio estable en un campo puramente electrostático. Esta situación es diferente para el caso de un campo magnetostático y en consecuencia la levitación magnética es posible, mientras que no lo es la electrostática. En este artículo, motivados por un problema análogo que aparece en sistemas mecánicos, mostramos que la adición de un pequeño campo eléctrico oscilante a una configuración electrostática conduce a la estabilización de los puntos de equilibrio inestables. De esta forma, la levitación resulta posible en ese campo "casi" electrostático.

The picture of a superconductor floating over a magnet is well known and is good proof that magnetic levitation is possible. Neither Maxwell equations nor potential theory are needed to convince someone about the feasibility of the phenomenon. The situation is quite different with electrostatic levitation. One can prove it is not possible, but some considerations from potential theory should be done [1]; in old textbooks [2] the demonstration, pompously called Earnshaw's theorem, establishes the non existence of stable equilibrium points in a purely electrostatic field. In this paper we show that a completely different physical problem could give us a clue to override the impossibility of levitation in an (almost) electrostatic field. The level of the article is adequated for a student that has taken a first course in Electrodynamics, and it may be used as an introduction to a subject of clear technological interest.

For mechanical systems it is well know that an inverted pendulum cannot be in a stable equilibrium, but the pendulum can be indefinitely in that position if a small oscillatory motion is applied to the support point. It is a typical exercise in control theory courses to find out the way to stabilise an inverted pendulum. Motivated by this result, this article shows that a small oscillating field superimposed on an electrostatic field can stabilise a charged particle that otherwise would be in unstable equilibrium.

The structure of this paper is as follows. In section I the results concerning the existence (or not) of stable equilibrium in magnetostatics (or electrostatics) are briefly reviewed. In section II the stabilising effect of an oscillating electric field on a charge particle in unstable equilibrium is proven.

## 1. Equilibrium in electrostatic and magnetostatic fields

Let us imagine a particle in a static field **F**(**r**). If this particle is in equilibrium at **r**$_0$, the net force on it should be zero:

$$\mathbf{F}(\mathbf{r}_0) = 0 \qquad (1)$$

However, this condition does not guarantee that the equilibrium is stable. To get such a condition, the particle should return to its original position if it is displaced from the equilibrium point, i.e. the force should push the particle back to the initial condition. In mathematical terms, this means that:

$$\nabla \mathbf{F}(\mathbf{r}_0) < 0 \tag{2}$$

The above conditions needed for the existence of a stable equilibrium are general. If the force can be derived from a potential $\psi$, i.e. $\mathbf{F} = -\nabla \psi$ equations (1) and (2) can be rewritten as:

$$\begin{aligned}\nabla \psi(\mathbf{r}_0) &= 0 \\ \nabla^2 \psi(\mathbf{r}_0) &> 0\end{aligned} \tag{3}$$

A rigorous proof of this can be found in the well-known book of Kellog about potential theory [3].

If we consider an electrostatic field $\varphi$, it is described in empty space by Laplace's equation:

$$\nabla^2 \varphi = 0 \tag{4}$$

Therefore, there is no way that conditions (3) are fulfilled at any point and there is no stable equilibrium point in a purely electrostatic field for a test charged particle.

Now let us consider the case of a small dielectric body in an electrostatic field. In this case a polarisation $\mathbf{P}$ is induced in the body by the electric field $\mathbf{E}$, and both quantities are related to each other by:

$$\mathbf{P} = \chi_e \mathbf{E} \tag{5}$$

where $\chi_e$ is the electrical susceptibility.

The polarisation **P** produces a dipolar moment **p** in the body. If the volume $V$ is small enough so that the electric field can be taken as a constant inside the body, then we can write:

$$\mathbf{p} = \int_V \mathbf{P} dV = \int_V \chi_e \mathbf{E} dV \cong \chi_e V \mathbf{E} \tag{6}$$

The force acting on a charged body in an electrostatic field is given by [2, 4]:

$$\mathbf{F}_e = (\mathbf{p} \cdot \nabla)\mathbf{E} \tag{7}$$

Using (6) and a well-known vectorial identity, eq. (7) can be rewritten as:

$$\mathbf{F}_e = \frac{1}{2} \chi_e V \nabla |\mathbf{E}|^2 \tag{8}$$

This last equation can be restated for the case of a small magnetic body in a magnetostatic field **H**. In this case, eq. (8) becomes:

$$\mathbf{F}_m = \frac{1}{2} \chi_m V \nabla |\mathbf{H}|^2 \tag{9}$$

where $\chi_m$ is the magnetic susceptibility.

From eqs. (8) and (9) we can understand why there is magnetic levitation but not an electrostatic one. According to (3), a stable equilibrium point requires that the force on the body to be zero at that point and the divergence of the force to be negative. In the present case, this means that:

$$\begin{aligned} \chi_e \nabla^2 |\mathbf{E}|^2 &< 0 \\ \chi_m \nabla^2 |\mathbf{H}|^2 &< 0 \end{aligned} \tag{10}$$

However, the Laplacian of a positive defined quantity is always positive; therefore the only way to fulfill (10) is with a negative susceptibility. But all known substances have positive electric susceptibility, and consequently electrostatic levitation is not feasible. Certainly there are materials with negative magnetic susceptibilities –the diamagnetic ones and superconductors- and for that reason magnetic levitation is possible.

The next section will show that the addition of a small oscillating electrical field to an electrostatic one allows the existence of stable equilibrium points. Thus levitation becomes, at least in principle, possible in an almost electrostatic field.

## 2. Stabilisation through a small oscillating field

In this section we analyse what happens when a small oscillating field is added to an electrostatic configuration. It turns out that the additional field can stabilise a charged particle, and consequently, electrostatic levitation becomes feasible.

Let us assume an electrostatic field $\mathbf{E}_0(\mathbf{r})$ and an unstable equilibrium point at $\mathbf{r}_0$; i.e. if a test particle with charge $q$ and mass $m$ is placed at $\mathbf{r}_0$, then:

$$\begin{aligned} q\mathbf{E}_0(\mathbf{r}_0) &= 0 \\ q\nabla.\mathbf{E}_0(\mathbf{r}_0) &> 0 \end{aligned} \tag{11}$$

A small oscillating field $\mathbf{E}_w(\mathbf{r}, t)$ is added. The particle position is now written as:

$$\mathbf{r} = \mathbf{r}_0 + \mathbf{f} \tag{12}$$

Since the added field is small and rapidly varying, it is assumed that $\mathbf{f}$ is also small and oscillates fast. It is also supposed that the fast changing vector $\mathbf{f}$ is controlled by the oscillating field $\mathbf{E}_w(\mathbf{r}, t)$ while the static component of the position vector is ruled by the electrostatic field $\mathbf{E}_0(\mathbf{r})$. For this reason, the equation of motion is written as follows:

$$\begin{aligned} m\ddot{\mathbf{f}} &= q\mathbf{E}_w(\mathbf{r},t) \\ &\cong q\mathbf{E}_w(\mathbf{r}_0,t) \end{aligned} \tag{13}$$

To write the second line of (13), it has been taken into account that $|\mathbf{f}| \ll |\mathbf{r}_0|$.

To solve the above equation, an explicit expression for the oscillating field should be chosen. It is assumed that the field is described by:

$$\mathbf{E}_w(\mathbf{r},t) = \mathbf{A}(\mathbf{r})e^{iwt} \tag{14}$$

**A** is the amplitude of the field and may change with position.

Given (14), equation (13) can be solved and the result is:

$$\mathbf{f} = -\frac{q}{mw^2}\mathbf{E}_w(\mathbf{r}_0,t) \tag{15}$$

Actually, this is an approximate solution of (13) since the field at $\mathbf{r}_0$ has been taken. We may improve the solution by expanding the field around $\mathbf{r}_0$, resulting in:

$$m\ddot{\mathbf{f}} = q\mathbf{E}_w(\mathbf{r}_0,t) + \mathbf{f}.\nabla\mathbf{E}_w(\mathbf{r}_0,t) + ... \tag{16}$$

To get a physically significant result, the magnitudes should be averaged over a complete period of the oscillating field. This temporal average will be denoted with brackets <...>. It is clear that the first term of the right side vanishes, and using (15) as an approximate value for **f**, it turns out that:

$$\begin{aligned} <m\ddot{\mathbf{f}}> &= \langle q\mathbf{E}_w(\mathbf{r}_0,t)\rangle + \langle \mathbf{f}.\nabla\mathbf{E}_w(\mathbf{r}_0,t)\rangle \\ &= -\frac{q}{mw^2}\langle \mathbf{E}_w(\mathbf{r}_0,t).\nabla\mathbf{E}_w(\mathbf{r}_0,t)\rangle \\ &= -\frac{q}{2mw^2}\langle \nabla|\mathbf{E}_w(\mathbf{r}_0,t)|^2\rangle \\ &= -\frac{q}{4mw^2}\nabla|\mathbf{A}(\mathbf{r}_0)|^2 \end{aligned} \tag{17}$$

Notice that an additional force $\mathbf{F}_w = <m\ddot{\mathbf{f}}>$ acts on the particle; however, the amplitude of the oscillating field may be chosen in such a way that this new force becomes zero at $\mathbf{r}_0$. What really matters is the divergence of $\mathbf{F}_w$:

$$\nabla \cdot \mathbf{F}_w = -\frac{q}{4mw^2} \nabla^2 |\mathbf{A}(\mathbf{r}_0)|^2 < 0 \qquad (18)$$

As we have seen above, the Laplacian of a positive function is always greater than zero; consequently the divergence of this extra force is negative. A fine-tuning of the oscillating field amplitude is needed to assure that the net force on the particle is zero, but the net divergence is negative. From (11), (17) and (18), we conclude that:

$$\begin{aligned} \nabla |\mathbf{A}(\mathbf{r}_0)|^2 &= 0 \\ \nabla^2 |\mathbf{E}(\mathbf{r}_0)|^2 - \frac{1}{4mw^2} \nabla^2 |\mathbf{A}(\mathbf{r}_0)|^2 &< 0 \end{aligned} \qquad (19)$$

These are the conditions $\mathbf{A}(\mathbf{r})$ should satisfy. No physical law is violated; they only require that the module of $\mathbf{A}(\mathbf{r})$ reaches an extreme local value at $\mathbf{r}_0$ with certain constraint related to the value of its Laplacian at that point. Once these conditions are fulfilled, the point $\mathbf{r}_0$ becomes a stable equilibrium point in an "almost electrostatic" field. This situation is completely analogous to the inverted pendulum, initially in unstable equilibrium, that is stabilised by an oscillating motion applied to the pivot point.

The purist may argue that the field is not more electrostatic, and that is certainly true. In this sense, Earnshaw's theorem is still valid. However, since we have control on the field frequency, it may be chosen high enough as to fulfil the second line of (19) with a very small oscillating field amplitude. That is why we talk of an "almost electrostatic" field.

In summary, this exercise shows that a test charged particle can be stabilised in an electrostatic field with the help of a small, high frequency oscillating field. And thus, electrostatic levitation becomes possible. The demostration is simple enough as to be within the scope of an undergraduate student with an intermediate course in Electrodynamics.

**Acknowledgement**

The author is supported by the National Research Council (CONICET) of Argentina.